\documentclass[letterpaper,twocolumn,10pt]{article}
\usepackage{usenix2019_v3}

\usepackage{graphicx}
\usepackage{booktabs}
\usepackage{amsmath}
\usepackage{amssymb}
\usepackage{algorithm}
\usepackage{algorithmic}
\usepackage{xurl}

\newcommand{\sys}{AoiZora}

\definecolor{algcomment}{HTML}{2B5EA7}
\renewcommand{\algorithmiccomment}[1]{\hfill{\color{algcomment}$\triangleright$~#1}}

\hypersetup{pdftitle={AoiZora: Compiler-Agnostic, Placement-Aware Auto-Parallelization for Video Diffusion Transformers}, pdfauthor={Kaijian Wang, Yuanyuan Xu, Fanjiang Ye, Ye Cao, Jingwei Zuo, T. S. Eugene Ng, Yarong Mu, Yuke Wang}}

\begin{document}

\date{}

\title{\Large \bf \sys{}: Compiler-Agnostic, Placement-Aware Auto-Parallelization\\for Video Diffusion Transformers}

\author{
{\rm Kaijian Wang$^{1}$, Yuanyuan Xu$^{1}$, Fanjiang Ye$^{1}$, Ye Cao$^{1}$, Jingwei Zuo$^{1}$,}\\
{\rm T.\ S.\ Eugene Ng$^{1}$, Yarong Mu$^{2}$, Yuke Wang$^{1}$}\\[0.5em]
$^{1}$Rice University \quad $^{2}$Google
} %

\maketitle

\begin{abstract}
Video diffusion transformers (DiTs) process tens of thousands of spatio-temporal tokens per denoising step, so low-latency inference and training commonly spread them across several accelerators. Auto-parallelization systems price a sharding on its logical graph, but its cost depends on the program the compiler generates and, on direct networks such as TPU sub-slices, on which chip runs each logical rank; both are visible only after compilation, while the chip assignment must be fixed before it. We present \sys{}, a compiler-agnostic auto-parallelization framework built on two IRs, one for the program a partitioner shards and one for the program a compiler emits; it chooses the logical mesh, the layout of every tensor and operation, and the device assignment of an unsharded model by pricing its compiled programs, without executing any candidate. At the core of \sys{} is \emph{evidence}, a compiler-neutral record of a compiled program that leaves its placement open, so that one compilation can be priced under many placements by a calibrated model of the accelerator network and the placement can determine which sharding is selected. On TPU v5e, \sys{} reduces the denoising latency of Wan~2.1 and HunyuanVideo by 1.23$\times$ and 1.32$\times$ on average over expert configurations and Alpa, finding each plan within 10 minutes; it also speeds up training on TPU v5p by 1.11$\times$ on 4 chips and 1.21$\times$ on 8 chips on average and parallelizes PyTorch programs on GPUs with the same core.
\end{abstract}

\section{Introduction}
\label{sec:intro}

Video diffusion transformers (DiTs) such as Wan~\cite{wan}, HunyuanVideo~\cite{hunyuanvideo}, and CogVideoX~\cite{cogvideox} run a denoiser tens of times per request over tens of thousands of spatio-temporal tokens, so low-latency inference and training spread them across several accelerators. Unlike LLMs, which have converged on a few well-known parallelization recipes~\cite{megatron}, DiTs are activation-bound, and their parallelization strategies mix context, head, tensor, guidance, and weight sharding~\cite{xdit,swiftfusion} in proportions that change with the model, resolution, frame count, and allocation (\S\ref{sec:eval-inference}). Auto-parallelization systems~\cite{alpa,nnscaler} are the natural answer, but they price the \emph{logical} graph, FLOPs and redistribution bytes on an abstract device mesh, a poor guide to what runs (\S\ref{sec:eval-ablation}).

\begin{figure}[t]
  \centering
  \includegraphics{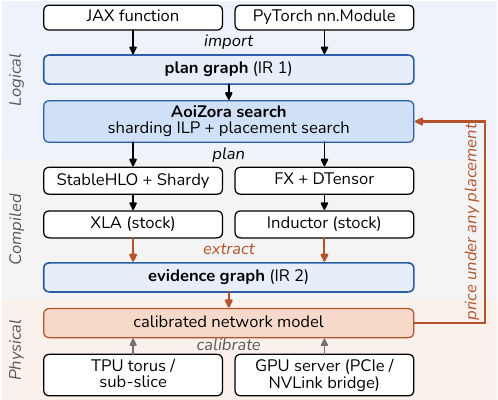}%
  \caption{\sys{} overview; prior systems see only the logical layer.}
  \label{fig:overview}
\end{figure}

Two layers below the logical graph decide the cost of a \emph{parallelization plan}, which consists of a logical mesh, a sharding of every tensor and operation over it, and an assignment of its ranks to chips (Figure~\ref{fig:overview}). \emph{First, the cost depends on the program the compiler generates, not on the logical graph these systems see:} partitioning, fusion, and scheduling decide how each redistribution is decomposed and tiled, which collectives overlap with compute, and how much memory is live, known only after compilation, in compiler-specific forms (\S\ref{sec:observations}).

\emph{Second, the cost depends on which physical chip runs each logical rank.} Frameworks treat each logical mesh axis as an independent channel~\cite{jax,devicemesh}, which holds inside a switch domain such as NVSwitch~\cite{h100}, but not on direct networks such as TPU tori and the \emph{sub-slices} that jobs actually receive~\cite{tpuv4,tpuv5e,tpuv5p}. Such direct networks train frontier models such as Gemini and Claude~\cite{gemini,anthropictpu,rainier}, and AWS Trainium2 follows the same design, linking the 16 chips of each instance in a 2D torus~\cite{trn2arch}. There, logical axes share links whenever the mesh does not factor onto physical dimensions (Figure~\ref{fig:placement}): one Wan~2.1 sharding on v5e-8 runs 16.6\% slower under its worst order of mesh axes than its best (\S\ref{sec:observations}).

Existing systems each see one layer. Auto-parallelization systems~\cite{alpa,nnscaler,galvatron} price the logical graph and leave placement to the framework's default mesh; network-aware placement methods~\cite{nest} assign stages to network levels from profiled costs but neither see the compiled program nor choose the chip of each rank; compilers see the program they emit but neither model the physical network nor choose placements; and topology mapping and collective synthesis~\cite{hoefler,taccl,tacos} optimize for a \emph{given} communication pattern but cannot change it. No system decides sharding and placement on the program that actually runs.

Our observation is that compilation already reveals both missing layers, if its output is captured in the right form: placement changes only the device groups of a compiled program (\S\ref{sec:observations}), so recording every compiled collective over \emph{logical ranks}, with the device assignment kept separate, makes one compilation priceable under any placement by a physical network model, and usable by a search that knows nothing about the compiler. We call this record its \emph{evidence}: what the compiled program computes and communicates, in what order, and how much memory it keeps live, independent of the compiler. We present \sys{}, a compiler-agnostic auto-parallelization framework that chooses the logical mesh, the layout of every value and operation, and the device assignment for DiT inference and training from compiled programs, without executing any candidate. Each piece of information this requires is known by a different party: the framework knows which plans are legal, the compiler what will run, the hardware what it costs, and the executable whether the plan was applied. We define an IR at each of the two points every graph compiler exposes, a plan graph for the program it partitions and an evidence graph for the program it emits (Figure~\ref{fig:overview}), and give each party one contract over them (\S\ref{sec:overview}). We use no recipes or DiT-specific knowledge and apply the selected plan through the stock compilation path; our implementation supports JAX/XLA and PyTorch/Inductor with one search core. We pin each plan so that the compiler cannot change it and extract evidence where the schedule is final (\S\ref{sec:evidence}), price evidence with a network model calibrated per slice that captures shared links (\S\ref{sec:network}), and search placements by rebinding the evidence of a pool of compiled shardings, so that placement can change which sharding is selected (\S\ref{sec:search}).

On TPU v5e, \sys{} reduces the denoising latency of Wan~2.1 and HunyuanVideo by 1.23$\times$ and 1.32$\times$ on average over the faster of the community default and Alpa (\S\ref{sec:eval-inference}), finding each plan within 10 minutes (\S\ref{sec:eval-planning}). It speeds up training by 1.11$\times$ and 1.21$\times$ on 4 and 8 chips on average over the faster of the official layout and Alpa on TPU v5p (\S\ref{sec:eval-training}) and, with the same core, parallelizes PyTorch programs on GPUs (\S\ref{sec:eval-portability}).

Our contributions include:
\begin{itemize}
  \item We present \sys{}, a compiler-agnostic auto-parallelization framework that jointly selects a DiT's sharding and placement by pricing compiled programs, not the logical graph, without running any candidate.
  \item We define two IRs: a plan graph into which JAX and PyTorch programs are imported, and evidence, a record of the compiled program whose collectives name logical ranks instead of chips, so that one compilation is priced under any placement by a network model that accounts for shared links.
  \item We implement \sys{} for JAX/XLA and PyTorch/Inductor with one shared search core, and apply each plan through the stock compiler.
  \item We show speedups of 1.23$\times$ and 1.32$\times$ on Wan~2.1 and HunyuanVideo inference on TPU v5e, and 1.11$\times$ and 1.21$\times$ on training on 4 and 8 chips of TPU v5p, over the faster of the community default and Alpa.
\end{itemize}

\section{Background and Motivation}
\label{sec:background}

\subsection{Background}
\label{sec:bg}

\textbf{DiT parallelism.} A DiT denoiser applies transformer blocks, each with self-attention over all spatio-temporal tokens, cross-attention to text tokens, and an MLP, to a latent whose activations rather than weights dominate memory and communication. Parallelization strategies therefore combine batch and classifier-free-guidance~\cite{cfg} parallelism, context parallelism over tokens~\cite{ringattention,ulysses}, head and tensor parallelism~\cite{ulysses,megatron}, and weight sharding gathered before use~\cite{zero,fsdp}, in proportions DiT systems tune per model~\cite{xdit,swiftfusion,maxdiffusion}. Shapes are static, so every step runs the same program.

\textbf{Direct networks and sub-slices.} TPU chips link directly to their neighbors without a switch~\cite{tpuv4,tpuv5e}. A v5e chip has four inter-chip links, and v5e forms a 2D torus~\cite{tpuv5e}; a v5p chip has two TensorCores, and v5p forms a 3D torus~\cite{tpuv5p}. A job receives a sub-slice without wrap-around links: a v5e-16 sub-slice is a $4{\times}4$ grid of chips, and a v5p-16 sub-slice, with 16 TensorCores, is a $2{\times}2{\times}2$ cube of eight chips~\cite{tpuv5p}. A device assignment maps the logical ranks of a mesh to chips.

\begin{figure}[t]
  \centering
  \includegraphics{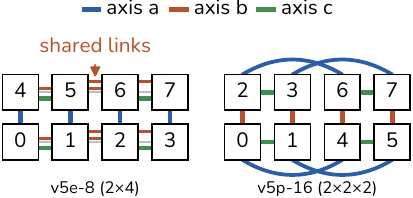}%
  \caption{A $2{\times}2{\times}2$ mesh on v5e-8 (left) strides one axis, sharing links; on v5p (right) every axis is a physical dimension.}
  \label{fig:placement}
\end{figure}

\textbf{Graph compilation.} Both pipelines in Figure~\ref{fig:overview} have the same shape. A partitioner applies a sharding to a whole-program graph: Shardy~\cite{shardy} on StableHLO~\cite{stablehlo} from JAX~\cite{jax}, or DTensor~\cite{dtensor,devicemesh} on an ATen graph from PyTorch~\cite{pytorch}. A compiler, XLA~\cite{gspmd,xla} or Inductor~\cite{inductor}, then fuses and schedules it into a program where collectives, their groups, and their overlap become concrete.

\subsection{Motivating Observations}
\label{sec:observations}

We characterize what decides the cost of a parallelization plan for a DiT on a TPU sub-slice with the stock JAX and XLA stack, without automatic search; each observation leads to one design of \sys{}.

\begin{figure}[t]
  \centering
  \includegraphics{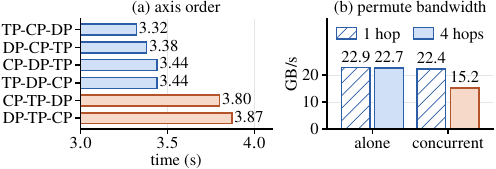}%
  \caption{(a) Time of one denoising step of the sharding of Observation~2, with 2-way context (CP), tensor (TP), and data (DP) parallelism, under each order of its axes on the chips. (b) Bandwidth of a permute route over 1 and 4 hops, alone and beside concurrent permutes that share links only with the 4-hop one.}
  \label{fig:motivation}
\end{figure}

\textbf{Observation 1: the compiler, not the sharding, decides the communication.} A sharding fixes the layout of every value; the compiler decides how values move between layouts. For Wan~2.1 at 480p and 81 frames, sharded 2-way over data, context, and tensor axes by MaxDiffusion~\cite{maxdiffusion}, JAX hands XLA a program with 48 sharding annotations and no collective. The partitioner turns it into 36 synchronous collectives per transformer block, 1,459 per step. Optimization then leaves 29 per block, 1,177 per step, in a different mix: it removes two of the three identical all-gathers that feed the query, key, and value projections, pairs small all-reduces into combined ones, fuses each all-reduce whose result is sliced into a reduce-scatter, and makes 18 of the 46 distinct collectives asynchronous so that they overlap with compute. None of this exists before compilation, and each compiler records it in its own form. Sharding annotations are also hints rather than commands: here they fix the entry parameters and two values, the partitioner decides every other layout, and the final executable can differ from earlier graphs, as when an asynchronous all-gather becomes synchronous and its overlap disappears. \emph{A plan must therefore be priced by the program the compiler emits, read where its schedule is final, and pinned so that the program that runs is the one priced (\S\ref{sec:evidence}).}

\textbf{Observation 2: one program costs differently under different placements.} We fix one sharding of Wan~2.1 14B at 480p and 81 frames on v5e-8, with 2-way context, tensor, and data parallelism, and run it under all six orders in which its three logical axes can be laid onto the chips. The computation, tensor shapes, and logical communication volume are identical, yet one denoising step takes 3.32 to 3.87\,s, a 16.6\% spread (Figure~\ref{fig:motivation}a). Hop count is not the cause: on the same slice, a permute route reaches 22.7 to 22.9\,GB/s over one or four hops and keeps 22.4\,GB/s beside concurrent traffic on other links, but drops to 15.2\,GB/s when that traffic shares its links (Figure~\ref{fig:motivation}b). A $2{\times}4$ grid cannot make three 2-way axes physical dimensions, so one axis is strided and its groups cross the links of another (Figure~\ref{fig:placement}); on the v5p cube, every axis is a physical dimension. Frameworks fix the placement when the mesh is built, by device order or a fixed rule, before the program that reveals this traffic exists. \emph{Placement is therefore part of the plan, and pricing it requires a network model of routes and link sharing (\S\ref{sec:network}).}

\textbf{Observation 3: placement changes only the device groups of the compiled program.} Compiling one Wan~2.1 sharding under two placements of a v5e-16 sub-slice yields the same program: the same fusions, the same schedule, and the same 64.7\,MB all-reduce on the tensor axis, whose replica groups alone differ, $\{0,1\},\{2,3\},\dots$ in one and $\{0,2\},\{4,6\},\dots$ in the other. Compiling every pair of sharding and placement is nevertheless impractical: a $4{\times}4$ logical mesh can be laid onto 16 chips in $16!/(4!\,4!) \approx 3.6{\times}10^{10}$ ways that differ in their device groups, and one compilation of the full Wan~2.1 denoiser takes 5.4 to 7.4\,s on v5e-8. Because shardings emit different collectives, a placement that suits one sharding can penalize another. \emph{Recording groups over logical ranks, with the assignment kept apart, lets one compilation be priced under every placement (\S\ref{sec:evidence}), and placement should be chosen jointly with sharding over a compiled pool (\S\ref{sec:search}).}

\section{Design Overview}
\label{sec:overview}

\textbf{Problem.} For an unsharded program $G$ on an allocation of $N$ devices, we choose a logical mesh $m$, legal layouts and strategies $L \in \mathcal{L}_m$ for every value and operation, and a device assignment $a$:
\begin{equation}
\min_{m,\,L,\,a}\ T_R(E_{m,L,a}) \quad \text{s.t.} \quad M(E_{m,L,a}) \le M_{\max},
\label{eq:problem}
\end{equation}
where $E_{m,L,a}$ is the evidence of $G$ compiled under the plan, $T_R$ the step time a resource model $R$ predicts from it, and $M$ the compiled memory per device. The search has a budget of time and compilations and executes no candidate.

$T_R$ exists only after compilation (Observation~1), while $a$ is a compilation input whose effect appears only in the device groups of the output (Observations~2 and~3). We rely on one approximation: if collective groups are recorded over logical ranks, then $E_{m,L,a'} \approx \textsc{Rebind}(E_{m,L,a}, a')$, and compiling under $a'$ confirms it. In our searches of Wan~2.1 and HunyuanVideo on v5e-8 and v5e-16, all 103 distinct plans recompiled under the selected assignment are priced exactly as their rebound evidence, so the approximation held in every case we observed.

\textbf{Two IRs.} We define a framework-neutral IR at each of the two points of \S\ref{sec:bg}. The \emph{plan graph}, at the partitioner's input, is a static single assignment (SSA)~\cite{ssa} graph over a closed set of operation kinds plus \emph{wrappers}, opaque operations such as attention that declare a sharding rule instead of a body; every legal layout and strategy derives from it. The \emph{evidence graph}, at the compiler's output, is evidence in concrete form (\S\ref{sec:evidence}). Both are analysis-only: a provenance map ties every original tensor and operation to the plan graph, a plan is applied by annotating the framework's program, and both assume a whole program with static shapes.

\textbf{Plan graph operations.} The plan graph has the following operation kinds, each with attributes and a shape contract that the graph validates: dot (with batch and contracting dimensions), convolution, elementwise, reduce, broadcast, reshape, transpose, slice, dynamic slice and dynamic update slice, slice scatter, split, concatenate, pad, reverse, constant, iota, empty, convert, and wrapper. Wrappers declare a sharding rule in factor notation; a factor present in an operand but missing from a result is reduced, and its axes are passed to the wrapper's local function. Attention is the wrapper \texttt{b h q d, b h k d, b h k d -> b h q d} with \texttt{k} and \texttt{d} replicated; its backward takes the forward output and log-sum-exp and reduces key and value gradients over the query factor.

\begin{figure*}[t]
  \centering
  \includegraphics{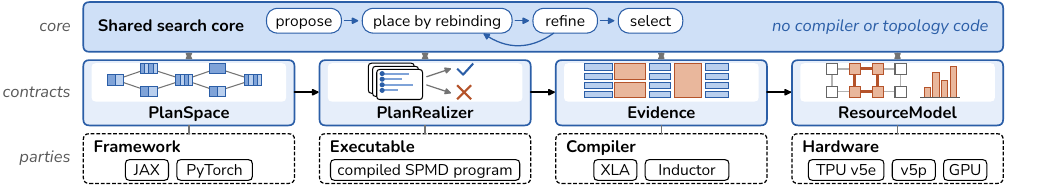}%
  \caption{The shared search core reaches each party only through its contract: the framework through a plan space (\S\ref{sec:search}), the executable through a realizer and the compiler through an evidence source (\S\ref{sec:evidence}), and the hardware through a resource model (\S\ref{sec:network}).}
  \label{fig:workflow}
\end{figure*}

\textbf{Contracts.} We give each party of \S\ref{sec:intro} one contract (Figure~\ref{fig:workflow}): the framework a plan space, the compiler an evidence source, the hardware a resource model, and the executable a realizer. They make the mechanisms of \S\ref{sec:evidence} to \S\ref{sec:search} independent of the compiler and the machine. The plan space has three operations. \emph{Propose} returns the layouts of complete candidates for a mesh, ranked by the integer program's own objective. \emph{Rebind} moves a candidate to another device assignment and keeps every logical decision; placement search uses it. \emph{Refine} proposes new candidates from compiled evidence; it may change their costs but never which layouts are legal or the memory limit. An evidence source returns an evidence graph or an explicit failure, and lists what it could not attribute instead of omitting it. A resource model prices proposals and evidence with the same constants. A realizer either produces an executable that implements the plan or fails; it never substitutes another plan. We keep the search, solver, placement search, and evaluator free of compiler and topology code; a new graph compiler needs an importer, a realizer, and an evidence adapter, and a new machine a resource model. We treat a plan that fails validation after planning as a bug, not a cue to fall back, and reuse a plan for every step with the same shapes.

\textbf{Planning flow.} A plan passes through the contracts in a fixed order. The plan space proposes a candidate for a mesh; the realizer pins and compiles it, and the evidence source records the compiled program over logical ranks; the resource model prices that evidence under the candidate's assignment and, by rebinding, under any other; placement search and refinement add candidates from these prices; and the selected candidate is the one with the lowest compiled prediction that fits in memory. Only proposal ranking uses the integer program's own estimate.

\section{Compiled Evidence and Realization}
\label{sec:evidence}

\textbf{Evidence graph.} Each event of the evidence graph has a kind (compute, matmul, attention, or a collective), a work shape and type, local bytes and flops or a tiled collective payload, groups of participants, and data, schedule, and asynchronous dependencies; the graph adds the assignment of logical ranks to devices, the per-device memory peak, and every item it could not attribute. Three rules make it safe to price. We record collective groups as logical ranks and store the assignment separately, so that one compilation is priced under any assignment (Observation~3). We mark what cannot be attributed as an explicit unknown, and reject a candidate with unattributed communication rather than credit it with free communication. We represent a collective whole or by its phases, never both.

\begin{figure}[t]
  \centering
  \includegraphics{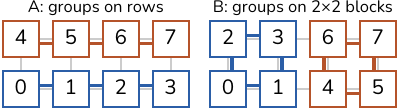}%
  \caption{One compiled all-gather with groups $\{0,1,2,3\}$ and $\{4,5,6,7\}$ over logical ranks, rebound to two assignments of a $2{\times}4$ sub-slice and priced from the same evidence.}
  \label{fig:rebinding}
\end{figure}

Rebinding to another assignment is a substitution (Figure~\ref{fig:rebinding}). It is exact when the compiler's output depends on the assignment only through the identities of group members, and is confirmed by compilation when the compiler may choose algorithms, layouts, or schedules from the topology. We therefore use rebound prices only to rank assignments and compile an assignment before selecting it; candidates with identical physical groups share evidence.

\textbf{Realizing a plan as priced.} Because compilers treat shardings as hints (Observation~1), we pin the whole plan through the provenance map, so the partitioner cannot propagate another layout through a value the plan leaves free. On JAX, the realizer constrains every value and consumer operand to the plan's layouts and compiles for the plan's device order, so the partitioner can insert only the redistributions the plan priced; on PyTorch, it runs each node over DTensors with operands and results redistributed to the plan's layouts.

We accept a candidate whose requested properties are satisfied or legally changed by the compiler, in fusion, collective decomposition, layouts, or scheduling, which the evidence then prices; we reject one whose device order or pinned layouts differ from the plan, or whose compiled artifacts cannot be matched to the executable, rather than misprice it.

\textbf{Extracting evidence.} We take evidence at the latest point at which collectives still carry their groups and the schedule is final: earlier, neither is decided; later, the compiler has merged collectives or lowered them into opaque calls. From XLA, this is the HLO module just before asynchronous collectives are fused; compute comes from XLA's per-instruction cost analysis, attention is attributed by operand shape, and collectives that the final executable runs synchronously are matched by channel, group, type, and loop count and charged at completion. From Inductor, it is the final kernel schedule, whose dependencies follow CUDA stream semantics. Memory comes from the final executable.

Five checks tie both IRs to the original program. \emph{Accounting}: every original tensor maps to a plan-graph value with identical shape and type, and every plan-graph operation has an origin. \emph{Semantics}: a NumPy~\cite{numpy} interpreter of the plan graph reproduces the JAX program's output and every intermediate value of the captured PyTorch graph. \emph{Wrappers}: a wrapper's local execution plus its declared reductions equals the global operation. \emph{Realization}: realized plans of randomly drawn integer-program solutions on all candidate meshes reproduce the eager program. \emph{Training}: the captured training step equals the framework's own autograd and optimizer step over all outputs.

\section{Pricing Evidence}
\label{sec:network}

We price evidence outside the compiler with a resource model that sees routes and link sharing (Observation~2) and target ranking fidelity: an error common to all candidates is harmless; one that swaps two of them picks the wrong plan.

\textbf{Shared evaluator.} We split pricing in two: a resource model prices each event in isolation and declares the share of every resource the event occupies while it runs, and a shared evaluator, which knows nothing about the hardware, schedules the evidence with those prices. Because a resource is either a link or a shared capacity pool, the same evaluator applies to direct networks and to switched networks, where flows share pools such as a PCIe switch or a host bridge; our GPU model is the second instance. We follow the compiler's schedule rather than an ideal one: an event starts when its dependencies, including asynchronous starts and dones, have finished, and we never reorder events to create overlap that the executable does not have.

\textbf{Compute.} Compute follows the roofline model~\cite{roofline}, $t_{\mathrm{op}} = \max(F/P\eta_c,\ B/W\eta_m,\ t_{\min})$, and attention takes $t_{\mathrm{attn}} = 4bhqkd/P\eta_a(k)$, for an operation with $F$ flops and $B$ local bytes on a chip with peak rate $P$ and memory bandwidth $W$, where $\eta_c$ and $\eta_m$ are fitted efficiencies and $t_{\min}$ is a minimum kernel time. Attention over $b$ batch rows, $h$ heads, $q$ queries, $k$ keys, and head dimension $d$ uses a fitted $\eta_a(k) = \eta_{\max} k/(k + k_0)$.

\textbf{Collectives.} Chips sit at their physical coordinates; neighbors share a full-duplex link of two directed channels, and a dimension wraps around only if the allocation spans it. A collective's template, chosen per kind and group geometry, decomposes it into phases: unidirectional, bidirectional, or per-dimension rings, whose members follow a Hamiltonian cycle of the group's grid when one exists, and direct flows for all-to-all and permutes. Each phase repeats a number of steps, and in each step its flows follow dimension-order routes, whose order is fitted per slice, through directed channels, physical links, and the injection and ejection of their end chips. The collective's isolated time is
\begin{equation}
T_c = t_0 + \sum_{p} s_p \Big(h_p\, t_{\mathrm{hop}} + \max_{r} \tau_{p,r}\Big),
\label{eq:collective}
\end{equation}
where phase $p$ repeats $s_p$ steps whose flows travel at most $h_p$ hops and route $B_{p,r}$ bytes through resource $r$, which needs $\tau_{p,r} = B_{p,r}/c_r(m_p)$ per step at capacity $c_r(m_p) = \beta_r\,\eta_{\mathrm{kind}}\,m_p/(m_p + m_0)$ for chunk size $m_p$; $\beta_r$ is the raw capacity of a channel, link, or chip endpoint, and $t_0$, $t_{\mathrm{hop}}$, $\eta_{\mathrm{kind}}$, and $m_0$ are fitted constants. Groups of one collective that share a link add their loads, and bytes use the physical tiled layout.

\textbf{From prices to occupancy.} Pricing a collective returns, besides $T_c$, the fraction of that time during which it keeps each resource busy, $u_{c,r} = \min(1, \sum_{p} s_p \tau_{p,r} / T_c)$. Links are not the only shared resource: collectives also compete for the collective engine and, depending on their kind, for compute and memory bandwidth, so each kind adds fitted shares of these resources; on TPU v5e, two collectives of different kinds do not overlap even on disjoint groups, which a share of one on the engine expresses. A compute event occupies compute and memory bandwidth in proportion to the two roofline terms of its time. Only collective prices depend on the assignment, which keeps rebinding cheap.

\textbf{Fluid schedule.} The evaluator advances every running event at its isolated rate, slowed by its most oversubscribed resource: $dx_e/dt = 1/(T_e\,\sigma_e(t))$, with load $L_r(t) = \sum_{e \in \mathcal{A}(t)} u_{e,r}$ and slowdown $\sigma_e(t) = \max\big(1, \max_{r} L_r(t)/\eta_r(n_r(t))\big)$ over the resources $e$ uses. Here $\mathcal{A}(t)$ is the set of running events, those whose dependencies have finished, $T_e$ is the isolated time of event $e$ ($T_c$, $t_{\mathrm{op}}$, or $t_{\mathrm{attn}}$) and $u_{e,r}$ its share of resource $r$, $x_e$ is the fraction of $e$ completed, $n_r$ counts the running events that use $r$, and $\eta_r(n)$ is a fitted efficiency of a link shared by $n$ events, equal to one for a single user and for every other resource. Time advances to the next completion. Collectives interact only through shared resources.

\textbf{Proposition 1 (fluid schedule).} Let $\sigma_e(t)$ be the slowdown of the fluid schedule, so that $T_e\,\dot{x}_e = 1/\sigma_e$.
(a) If no other running event uses a resource that $e$ uses, then $\sigma_e = 1$ and $e$ progresses at its isolated rate. Hence isolated events, events on disjoint resources, and dependent chains are scheduled exactly as their isolated times compose along the dependencies. \emph{Proof.} Then $L_r = u_{e,r} \le 1$ and $\eta_r(1) = 1$ for every resource $r$ of $e$.
(b) No resource is driven beyond its capacity: $\sum_{e} u_{e,r}\, T_e\, \dot{x}_e \le \eta_r(n_r)$ for every resource $r$ and time $t$. \emph{Proof.} For every running $e$ that uses $r$, $T_e\,\dot{x}_e = 1/\sigma_e \le \eta_r(n_r)/L_r$, so the sum is at most $\eta_r(n_r)\, L_r^{-1} \sum_e u_{e,r} = \eta_r(n_r)$.
(c) If every $\eta_r$ is non-increasing in the number of users, starting another event never makes a running event faster, because it can only increase every $L_r$ and $n_r$.

\textbf{Calibration.} We price refined proposals (\S\ref{sec:search}) and evidence on physical routes with one set of constants, fitted once per chip type, slice, and compiler version on workload-independent microbenchmarks and never on the workload. On the allocation's own chips, they measure point-to-point transfers one to three hops apart, in one and both directions; every physical link in each direction, within and across hosts; all-gather, reduce-scatter, all-reduce, and all-to-all over every group geometry a plan can produce; one to four flows sharing one link; routing probes; two different collectives started together; a collective concurrent with a matrix product or a memory-bound update; matrix product, elementwise, and smallest-operation compute; and the attention kernel on one chip over batch-head products, query lengths, and key lengths. Each value is the slope of a device loop between two trip counts minus a loop of the same glue without the operation. On a multi-host slice every host runs the same loops. We never fit link sharing per scenario but predict it from routes and capacities alone, and test this by refitting the model without any link-sharing sample (\S\ref{sec:eval-ablation}). Disabling link sharing, sharing among concurrent events, or physical routes yields the evaluated variants.

\section{Searching Over Plans}
\label{sec:search}

\textbf{Proposing shardings.} We propose shardings with an operator-level integer program. For each logical mesh, every plan-graph value chooses a layout that assigns whole mesh axes to its dimensions, and every operation a strategy derived from its kind or wrapper rule; edges carry redistribution costs, and resident plus live memory must fit the device. The program is solved with HiGHS~\cite{highs} after the size reductions below. Pricing proposals on every assignment would couple the integer program to placement and multiply its solves by the number of assignments, while the assignment it scores is not the one the compiled program will prefer. We instead solve it once per mesh with every collective priced by the same templates and constants as evidence, but with each step crossing one hop on its own links: no routes and no sharing between groups. The program then depends on the mesh alone, and the assignment is left to the stages that see compiled programs.

\textbf{Integer program reductions.} To keep the program solvable on full models, \sys{} merges an operation whose strategies match its output's layouts one-to-one into that value, as Alpa does~\cite{alpa}; ties repeated regions of an unrolled program to one decision per layer position, which restricts corresponding occurrences to the same strategy; coalesces values bound by equality constraints; groups identical rows and columns of pairwise cost factors while coupling their marginals; drops live-memory rows dominated by another; makes an auxiliary variable continuous when integral decisions determine it; and warm-starts each mesh with the embedded solutions of coarser meshes. The other reductions preserve the feasible assignments and the objective of the program they are applied to. Enumeration cuts exclude a found assignment through its integral decisions with more than one option only, which excludes exactly that assignment with a tighter relaxation.

\begin{algorithm}[t]
\caption{Placement by rebinding}
\label{alg:place}
\small\raggedright
\begin{algorithmic}[1]
\REQUIRE pool $C$ of a mesh $m$; axis-aligned assignments $A_0(m)$
\ENSURE assignment $a^\ast$ and the candidate of $C$ it selects
\FUNCTION[Full schedule]{\textsc{Cost}$(a)$}
\STATE \textbf{return} $\min_{c \in C} T_R(\textsc{Rebind}(E_c, a))$
\ENDFUNCTION
\STATE $a^\ast \gets \arg\min_{a \in A_0(m) \cup \{a_c\}_{c \in C}} \textsc{Cost}(a)$ \COMMENT{Starts}
\REPEAT
\STATE $a' \gets \arg\min_{a \in \mathcal{N}(a^\ast)} \min_{c \in C} T_{\mathrm{iso}}(\textsc{Rebind}(E_c, a))$
\SHORTIF{$\textsc{Cost}(a') < \textsc{Cost}(a^\ast)$}{$a^\ast \gets a'$}
\UNTIL{$a^\ast$ does not change or the time budget ends}
\STATE \textbf{return} $a^\ast$, $\arg\min_{c \in C} T_R(\textsc{Rebind}(E_c, a^\ast))$
\end{algorithmic}
\end{algorithm}

Three choices make it usable in the loop. First, we compile proposals in device-id order; placement search then chooses the assignment, and refinement re-solves the program on it with the same resource model as evidence, so refined proposals see routes and link sharing. We score every emitted program only from its compiled evidence. Second, we derive redistribution bytes from layout rules observed in compiled evidence, such as tile padding of gathered shards, not a default tiling. Third, after each solution we add a structural cut requiring the next to change at least one large tensor's layout, so the pool holds distinct plans.

\textbf{Placing by rebinding.} Algorithms~\ref{alg:place} and~\ref{alg:search} use the notation of Equation~\ref{eq:problem}. A compiled candidate $c$ carries its layouts $L_c$, assignment $a_c$, evidence $E_c$, and compiled prediction $T(c) = T_R(E_c)$; $\textsc{Compile}(m, L, a)$ realizes, compiles, and validates a plan and returns its candidate, or nothing if the plan is violated (\S\ref{sec:evidence}); $A_0(m)$ is the set of axis-aligned assignments, each giving every axis of mesh $m$ whole prime sub-axes of the slice's dimensions, the starts of Algorithm~\ref{alg:place}; $a_{\mathrm{id}}$ is device-id order; $\textsc{Rebind}(E, a)$ is evidence $E$ rebound to assignment $a$; and $T_{\mathrm{iso}}(E)$ sums the isolated times $T_c$ (Equation~\ref{eq:collective}) of the collectives in $E$. The compiled candidates of one mesh form a pool $C$, and placement search, $\textsc{Place}(C)$ in Algorithm~\ref{alg:place}, finds the assignment $a$ that minimizes $\textsc{Cost}(a)$, the best predicted step of any candidate in $C$ under $a$; it only rebinds evidence. We search a pool, not one candidate, so placement can choose the sharding (Observation~3). Assignments are too many to enumerate, so we refine structured starts locally (Algorithm~\ref{alg:place}). We screen neighbors by $T_{\mathrm{iso}}$, which needs no schedule, and judge only the best by the full schedule.

The starts $A_0(m)$ give every prime sub-axis of the grid to one logical axis, an owner map $o$, and place a rank by spreading the digits of each coordinate over the sub-axes its axis owns; on a $2{\times}4$ grid, a 4-way axis may form a line, a block, or a strided block. Assignments that induce the same device groups, up to renaming equal-size logical axes and transposing equal-size physical dimensions, are kept once. A neighbor in $\mathcal{N}(a)$ transposes two digits of the mixed-radix rank index, moving whole groups at once, or swaps two equal contiguous blocks of ranks. Starts cover regular product layouts; local moves reach irregular ones.

\textbf{Proposition 2 (structure of starts).} In a start with owner map $o$, the groups of logical axis $i$ are the sets of chips obtained by varying the digits of the sub-axes in $o^{-1}(i)$ over all their values while every other digit stays fixed. \emph{Proof.} Let $\ell$ be the logical coordinates of a rank. Spreading maps the digits of $\ell_i$ one-to-one onto the sub-axes in $o^{-1}(i)$, and no other coordinate of $\ell$ contributes to those sub-axes. Within a dimension, the sub-axes' factors and strides form a mixed-radix system, so the map from digits to physical coordinates is a bijection. Varying $\ell_i$ with the other logical coordinates fixed therefore varies exactly the digits of $o^{-1}(i)$.

\textbf{Corollary 1 (compact groups are among the starts).} If logical axis $i$ owns the minor sub-axes of one dimension, those with the smallest strides, its groups are contiguous segments along that dimension; if it owns the minor sub-axes of two dimensions, its groups are contiguous rectangles; otherwise they are strided lines or blocks. Hence, for every logical axis whose length is a product of minor prime factors of one or two dimensions, some start gives that axis contiguous groups, and Algorithm~\ref{alg:place} scores it before any local move. In the plans we evaluate, the groups of every all-gather, reduce-scatter, all-reduce, and all-to-all are the groups of a set of logical mesh axes, because the partitioner forms replica groups over mesh axes. The intuitive placements, which keep the groups of the most communication-heavy axis contiguous, are therefore always among the starts.

\textbf{Proposition 3 (ring steps).} If a group is a contiguous rectangle with an even side, consecutive members of its ring, including the last and the first, are physical neighbors, so every ring step crosses one hop. If it is a contiguous line without wrap-around, every step crosses one hop except the closing one, which crosses $n-1$ hops for $n$ members. If it is a strided line with stride $g$, every step crosses $g$ hops. \emph{Proof.} The ring order goes down the first column and back through the remaining columns in alternating directions, which is a Hamiltonian cycle of a grid with an even side; a line has no Hamiltonian cycle, so its order closes across the whole line; and consecutive members of a strided line are $g$ chips apart.

\textbf{Proposition 4 (number of starts).} Before deduplication, the construction of starts examines at most $d^{|S|}$ owner maps for a mesh of $d$ logical axes, where $|S| = \sum_j \Omega(n_j)$ counts the prime factors of the grid's dimensions with multiplicity, against $N!$ assignments. A $2{\times}4$ grid has $|S| = 3$, so a three-axis mesh has at most 27 starts against $8! = 40{,}320$ assignments; a $4{\times}4$ grid has $|S| = 4$, so at most 81 against $16! \approx 2.1\times10^{13}$. After deduplication, a $(2,4)$ mesh on a $2{\times}4$ grid has three starts, a line, a block, and a strided block, and $(2,8)$ and $(4,4)$ meshes on a $4{\times}4$ grid have two and three.

\textbf{Refining with evidence.} Rebinding can change the ranking of compiled candidates but cannot introduce shardings absent from the pool, so we solve the integer program again under the selected assignment (Algorithm~\ref{alg:search}), warm-started from the incumbent and priced on that assignment's routes, with the same legality and memory constraints and known plans excluded, so every refined proposal is a sharding the pool lacks. Placement search and refinement alternate over assignment and sharding. Because selection is over a pool that only grows, the best compiled prediction never increases, and every candidate the loop can select has been compiled on its own assignment. Alternation can still stop at a pair where neither stage alone improves: placement keeps the sharding, and refinement keeps the assignment. When a round does not improve a mesh's best compiled prediction by a fraction $\theta$ ($0.5\%$), we restart refinement under the start in $A_0(m)$ with the next lowest $\textsc{Cost}$, up to twice, before the mesh is exhausted. Because the memory term omits compiler temporaries, a mesh that compiled above the limit tightens its constraint by the ratio of compiled to estimated peak and is proposed again. The loop runs under a budget of time and compilations: the first compilations cover the best candidate of every mesh, and a compilation starts only if it is expected to finish in time.

\begin{algorithm}[t]
\caption{\sys{} planning loop}
\label{alg:search}
\small\raggedright
\begin{algorithmic}[1]
\REQUIRE plan graph $P$; logical meshes $\mathcal{M}$; memory limit $M_{\max}$; budget $B$
\ENSURE the selected candidate
\FORALL[Propose, unplaced]{$m \in \mathcal{M}$}
\STATE $C_m \gets \{\textsc{Compile}(m, L, a_{\mathrm{id}})\}_{L \in \textsc{Propose}(P, m)}$
\ENDFOR
\WHILE{$B$ remains and some mesh is not exhausted}
\FORALL{$m \in \mathcal{M}$ not exhausted}
\STATE $(a, c) \gets \textsc{Place}(C_m)$ \COMMENT{Algorithm~\ref{alg:place}}
\IF[Confirm]{$\textsc{Cost}(a) < (1-\theta) \min_{c' \in C_m} T(c')$}
\STATE $C_m \gets C_m \cup \{\textsc{Compile}(m, L_c, a)\}$
\ENDIF
\STATE $c \gets \arg\min_{c' \in C_m} T(c')$
\STATE $L' \gets \textsc{Refine}(P, m, c)$ \COMMENT{Under $a_c$}
\STATE $c'' \gets \textsc{Compile}(m, L', a_c)$; $C_m \gets C_m \cup \{c''\}$
\IF{$T(c'') \ge (1-\theta)\,T(c)$}
\STATE restart under the next start, or mark $m$ exhausted
\ENDIF
\ENDFOR
\ENDWHILE
\STATE \textbf{return} $\arg\min_{c \in \bigcup_m C_m,\, M(E_c) \le M_{\max}} T(c)$
\end{algorithmic}
\end{algorithm}

\section{Implementation}
\label{sec:impl}

\begin{figure*}[t]
  \centering
  \includegraphics{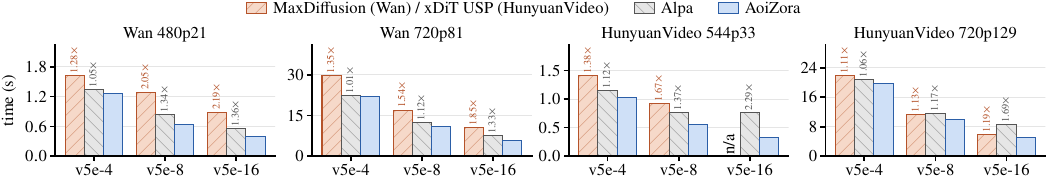}%
  \caption{Time of one denoising step for a short and a long video of each model on TPU v5e slices of 4, 8, and 16 chips (lower is better); each baseline bar is labeled with \sys{}'s speedup over it. N/a: USP cannot run a workload whose token count does not divide by 16.}
  \label{fig:inference}
\end{figure*}

\sys{} is a Python package of about 10K lines; we added the PyTorch backend without changing the core (\S\ref{sec:overview}).

\textbf{JAX.} The model is lowered under a replicated context that fixes only the device count, and StableHLO is imported one operation per plan-graph operation, keeping SSA names.

\textbf{PyTorch.} The model is captured with \texttt{torch.export} and decomposed to Core ATen, keeping linear layers, attention, and convolutions whole. We route linear layers, attention, and patch embedding above autograd through planner-visible operators, which keeps their sequence dimension shardable. Planning runs on fake tensors with a fake process group. We use PyTorch 2.14.1.

\textbf{Planning without hardware.} We plan for a TPU slice from its topology description on a CPU-only host, and the plan is byte-identical to one planned on the slice.

\section{Evaluation}
\label{sec:eval}

We evaluate end-to-end inference (\S\ref{sec:eval-inference}) and training (\S\ref{sec:eval-training}), planning cost (\S\ref{sec:eval-planning}), a second compiler stack (\S\ref{sec:eval-portability}), model fidelity and placement (\S\ref{sec:eval-ablation}), and numerical correctness (\S\ref{sec:eval-numerics}).

\subsection{Experimental Setup}
\label{sec:eval-setup}

\textbf{Testbed.} We run inference on TPU v5e-4, v5e-8, and v5e-16 ($2{\times}2$, $2{\times}4$, and $4{\times}4$ chips on 1, 1, and 4 hosts, 16\,GB HBM per chip) and training on TPU v5p-8 and v5p-16 ($2{\times}2{\times}1$ and $2{\times}2{\times}2$ chips on 1 and 2 hosts, 95\,GB per chip)~\cite{tpuv5e,tpuv5p}. We use JAX 0.11.2 and libtpu 0.0.48.

\textbf{Workloads.} We run inference of two open video DiTs with different block structures, Wan~2.1 14B~\cite{wan} at 480p and 720p with 21 to 81 frames and HunyuanVideo 13B~\cite{hunyuanvideo} at 544p and 720p with 33 to 129 frames, three lengths per resolution. The shapes range from short clips, where weights and per-layer collectives dominate a step, to long videos, where attention dominates. We train Wan~2.1 1.3B at the same shapes, with a global batch of two on 4 chips and four on 8 chips.

\textbf{Baselines.} For inference we compare with each model's community default, MaxDiffusion's default~\cite{maxdiffusion} for Wan and xDiT's unified sequence parallelism (USP)~\cite{usp,xdit} for HunyuanVideo, run as fixed plans on the same stack, and with Alpa~\cite{alpa}, run with its own resharding costs, strategies, and memory rows over a one-dimensional mesh and every two-dimensional view, with devices in id order; each setting reports Alpa's fastest valid plan. For training we compare with the official layout and Alpa.

\textbf{Methodology.} For inference we report the median time of one denoising step, and for training, of one training step; each median is over five synchronized steps after two warmup steps (three above 40,000 video tokens). \sys{} plans each workload within the slice's memory, and we run the plan it ranks first and check its outputs against an FP32 reference, or in training against the official layout's losses. \S\ref{sec:eval-ablation} reports the resource model's held-out error, 4.5\% to 6.0\% median on collectives, and \S\ref{sec:eval-numerics} the plans' numerical error, which stays within BF16 precision in inference and training. We report the geometric mean of \sys{}'s speedup over the faster baseline per setting.

\subsection{End-to-End Inference}
\label{sec:eval-inference}

Across all 36 settings, of which Figure~\ref{fig:inference} shows a short and a long video per model, \sys{} is the fastest, with a geometric-mean speedup over the faster baseline of 1.23$\times$ on Wan (1.01$\times$ to 1.71$\times$) and 1.32$\times$ on HunyuanVideo (1.06$\times$ to 2.29$\times$); Table~\ref{tab:app-inference} lists all 36 settings. Three effects explain these gains.

\begin{table*}[t]
  \centering
  \small
  \caption{Inference: seconds per denoising step on TPU v5e slices of 4, 8, and 16 chips. Default is MaxDiffusion (Wan) or xDiT USP (HunyuanVideo); n/a: USP cannot run 544p on 16 chips; speedup is \sys{}'s speedup over the faster baseline.}
  \label{tab:app-inference}
  \vskip 0.1in
  \setlength{\tabcolsep}{2pt}
  \begin{tabular*}{\textwidth}{@{\extracolsep{\fill}}llrrrrrrrrrrrr@{}}
    \toprule
    & & \multicolumn{4}{c}{4 chips} & \multicolumn{4}{c}{8 chips} & \multicolumn{4}{c}{16 chips} \\
    \cmidrule(lr){3-6} \cmidrule(lr){7-10} \cmidrule(l){11-14}
    Model & Workload & \sys{} & Default & Alpa & Speedup & \sys{} & Default & Alpa & Speedup & \sys{} & Default & Alpa & Speedup \\
    \midrule
    Wan 2.1 14B & 480p21 & 1.270 & 1.624 & 1.338 & 1.05$\times$ & 0.632 & 1.293 & 0.844 & 1.34$\times$ & 0.407 & 0.889 & 0.551 & 1.36$\times$ \\
     & 480p41 & 2.614 & 3.295 & 2.755 & 1.05$\times$ & 1.308 & 2.155 & 1.626 & 1.24$\times$ & 0.667 & 1.271 & 1.139 & 1.71$\times$ \\
     & 480p81 & 5.993 & 8.036 & 6.076 & 1.01$\times$ & 2.956 & 4.704 & 3.536 & 1.20$\times$ & 1.510 & 2.397 & 2.416 & 1.59$\times$ \\
     & 720p21 & 3.345 & 4.395 & 3.537 & 1.06$\times$ & 1.647 & 2.894 & 2.197 & 1.33$\times$ & 0.976 & 1.869 & 1.421 & 1.46$\times$ \\
     & 720p41 & 8.074 & 10.74 & 8.325 & 1.03$\times$ & 4.219 & 6.248 & 4.674 & 1.11$\times$ & 2.053 & 4.088 & 2.982 & 1.45$\times$ \\
     & 720p81 & 22.03 & 29.66 & 22.31 & 1.01$\times$ & 11.04 & 16.99 & 12.33 & 1.12$\times$ & 5.639 & 10.41 & 7.480 & 1.33$\times$ \\
    \midrule
    HunyuanVideo 13B & 544p33 & 1.024 & 1.408 & 1.149 & 1.12$\times$ & 0.553 & 0.924 & 0.760 & 1.37$\times$ & 0.335 & n/a & 0.767 & 2.29$\times$ \\
     & 544p65 & 2.524 & 3.198 & 2.832 & 1.12$\times$ & 1.304 & 1.762 & 1.719 & 1.32$\times$ & 0.832 & n/a & 1.627 & 1.96$\times$ \\
     & 544p129 & 7.247 & 8.485 & 7.713 & 1.06$\times$ & 3.747 & 4.504 & 4.567 & 1.20$\times$ & 2.194 & n/a & 3.847 & 1.75$\times$ \\
     & 720p33 & 2.216 & 2.850 & 2.523 & 1.14$\times$ & 1.160 & 1.609 & 1.505 & 1.30$\times$ & 0.595 & 1.083 & 1.484 & 1.82$\times$ \\
     & 720p65 & 6.389 & 7.470 & 7.021 & 1.10$\times$ & 3.257 & 3.985 & 3.974 & 1.22$\times$ & 1.677 & 2.245 & 3.415 & 1.34$\times$ \\
     & 720p129 & 19.70 & 21.88 & 20.91 & 1.06$\times$ & 9.974 & 11.26 & 11.63 & 1.13$\times$ & 5.038 & 5.993 & 8.530 & 1.19$\times$ \\
    \bottomrule
  \end{tabular*}
\end{table*}

\textbf{The gain grows with the slice because \sys{} changes the mesh.} MaxDiffusion's default places one context axis across the slice and shards every weight along it, so the bytes each chip gathers per layer stay nearly constant while its share of the computation shrinks; on Wan it scales by at most 3.35$\times$ from 4 to 16 chips. \sys{} instead selects two-axis meshes in 10 of the 12 Wan settings on 8 and 16 chips ($2{\times}4$, $2{\times}8$, and $4{\times}4$), which keep each collective within 2 to 8 chips, and scales at 78\% to 99\% of linear over the same range, so its speedup over the default rises from 1.31$\times$ on 4 chips to 1.90$\times$ on 16. On 4 chips, where it mostly keeps a one-dimensional mesh, it gains by choosing per tensor which operands to shard.

\textbf{The gain shrinks where attention dominates.} Quadratic attention takes a growing share of each step as videos lengthen and leaves less to layout. Over xDiT USP, the speedup falls from 1.38$\times$ to 1.17$\times$ at 544p on 4 chips. On 16 chips USP cannot run 544p, whose token counts do not divide by 16, while \sys{} uses a $4{\times}4$ mesh.

\textbf{Pricing the logical graph misses costs that compilation exposes.} Alpa is faster than the community default in 27 of the 33 settings where both run, yet slower than \sys{} in all 36, and the gap widens with the slice, from 1.04$\times$ and 1.10$\times$ on 4 chips to 1.48$\times$ and 2.02$\times$ on 16 chips (geometric means). Its integer program prices communication and resharding but not the differences in computation and memory traffic between strategies~\cite{alpa}, which compilation exposes (Observation~1), and keeps the devices in id order, so it cannot exploit placement (Observation~2).

\subsection{Training}
\label{sec:eval-training}

\begin{figure}[t]
  \centering
  \includegraphics{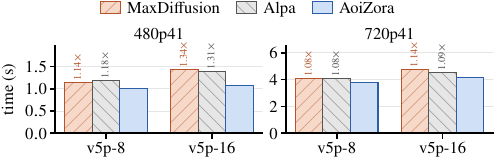}%
  \caption{Time of one training step of Wan~2.1 1.3B at two shapes on TPU v5p-8 and v5p-16 (lower is better); each baseline bar is labeled with \sys{}'s speedup over it.}
  \label{fig:training}
\end{figure}

Across all six shapes, of which Figure~\ref{fig:training} shows two, \sys{} trains Wan~2.1 1.3B faster than both baselines, with a geometric-mean speedup over the faster one of 1.11$\times$ on 4 chips and 1.21$\times$ on 8 chips (Table~\ref{tab:app-training} lists every shape).

\begin{table}[t]
  \centering
  \footnotesize
  \caption{Training of Wan~2.1 1.3B: seconds per training step on TPU v5p-8 (4 chips, global batch two) and v5p-16 (8 chips, global batch four). Default is MaxDiffusion's official training layout; gain is \sys{}'s speedup over the faster baseline.}
  \label{tab:app-training}
  \vskip 0.1in
  \setlength{\tabcolsep}{2pt}
  \begin{tabular*}{\columnwidth}{@{\extracolsep{\fill}}lrrrrrrrr@{}}
    \toprule
    & \multicolumn{4}{c}{v5p-8} & \multicolumn{4}{c}{v5p-16} \\
    \cmidrule(lr){2-5} \cmidrule(l){6-9}
    Shape & \sys{} & Default & Alpa & Gain & \sys{} & Default & Alpa & Gain \\
    \midrule
    480p21 & 0.449 & 0.546 & 0.586 & 1.21$\times$ & 0.491 & 0.682 & 0.698 & 1.39$\times$ \\
    480p41 & 1.000 & 1.137 & 1.185 & 1.14$\times$ & 1.065 & 1.424 & 1.391 & 1.31$\times$ \\
    480p81 & 2.717 & 2.887 & 2.931 & 1.06$\times$ & 2.867 & 3.344 & 3.231 & 1.13$\times$ \\
    720p21 & 1.393 & 1.621 & 1.588 & 1.14$\times$ & 1.458 & 1.972 & 1.881 & 1.29$\times$ \\
    720p41 & 3.762 & 4.055 & 4.060 & 1.08$\times$ & 4.164 & 4.728 & 4.536 & 1.09$\times$ \\
    720p81 & 11.82 & 12.15 & 12.30 & 1.03$\times$ & 12.29 & 14.01 & 13.45 & 1.09$\times$ \\
    \bottomrule
  \end{tabular*}
\end{table}

\textbf{The inference findings carry over to training.} The official layout, like MaxDiffusion's inference default, shards the sequence over all chips; \sys{} selects a $2{\times}4$ mesh on 8 chips and a one-dimensional mesh on 4 chips, where its gain shrinks from 1.21$\times$ on the shortest sequence to 1.03$\times$ on the longest as attention takes a growing share.

\textbf{\sys{} is faster than Alpa on every shape.} It is 1.04$\times$ to 1.42$\times$ faster than Alpa, which beats the official layout in one of six shapes on 4 chips and in five on 8 chips.

\subsection{Planning Cost}
\label{sec:eval-planning}

\textbf{Planning fits a fixed budget and is repaid within 139 videos.} \sys{} plans every inference workload within 240\,s on 4 and 8 chips and finds the plan it selects within 10 minutes on 16 chips. A plan serves every denoising step and request with its shape, so the time \sys{} saves per generated 30-step video over the faster baseline repays planning after 7 to 139 videos on 8 and 16 chips, fewest for long videos (720p129 and 720p81 on 8 chips) and most for short clips (480p21 on 16 chips).

\begin{table}[t]
  \centering
  \small
  \caption{Planning Wan~2.1 480p41 on v5e-8 as the search adds its stages; pairs are candidate and placement pairs priced by rebinding.}
  \label{tab:reuse}
  \vskip 0.1in
  \setlength{\tabcolsep}{2pt}
  \begin{tabular*}{\columnwidth}{@{\extracolsep{\fill}}lccc@{}}
    \toprule
     & Proposals & + placement & + refinement \\
    \midrule
    Pairs priced & 0 & 79 & 231 \\
    Programs compiled & 2 & 2 & 8 \\
    Proposal (s) & 73 & 73 & 73 \\
    Compilation (s) & 14 & 14 & 58 \\
    Placement (s) & 0 & 1.8 & 7.7 \\
    Refinement (s) & 0 & 0 & 92 \\
    Total (s) & 88 & 90 & 233 \\
    \bottomrule
  \end{tabular*}
\end{table}

\begin{table}[t]
  \centering
  \small
  \caption{Rebound and recompiled predictions and measured seconds per step of Wan~2.1 720p21 on v5e-8; the assignments place the four-way axis on $2{\times}2$ blocks, a line, or strided blocks (\S\ref{sec:search}).}
  \label{tab:rebind}
  \vskip 0.1in
  \setlength{\tabcolsep}{2pt}
  \begin{tabular*}{\columnwidth}{@{\extracolsep{\fill}}llccc@{}}
    \toprule
    Plan & Assignment & Rebound & Compiled & Measured \\
    \midrule
    Unrefined & block (searched) & 1.7934 & 1.7934 & 1.6540 \\
     & line & 1.8870 & 1.8870 & 1.7792 \\
     & strided block & 1.8872 & 1.8872 & 1.7435 \\
    \midrule
    Refined & block (searched) & 1.7730 & 1.7730 & 1.6474 \\
     & line & 1.8939 & 1.8939 & 1.7855 \\
     & strided block & 1.8941 & 1.8941 & 1.7447 \\
    \bottomrule
  \end{tabular*}
\end{table}

\textbf{Rebinding makes placement search affordable and exact.} For Wan 480p41 on v5e-8, the full search prices 231 candidate and placement pairs by rebinding in 7.7\,s, 18 to 46\,ms each, and compiles eight programs (Table~\ref{tab:reuse}); compiling every priced pair instead, at a median of 5.5\,s each, would take an estimated 21 minutes, about 165 times longer. Rebinding loses nothing. For both the refined and unrefined plans of Wan 720p21 on v5e-8 (both on a $2{\times}4$ mesh), the evidence compiled on the searched assignment and rebound to each other assignment receives exactly the same price as the program compiled on that assignment, in every pair (Table~\ref{tab:rebind}), as Observation~3 expects. The model ranks the searched assignment first. The line and strided-block assignments measure 5.4\% to 8.4\% slower, which the model predicts as 5.2\% to 6.8\%.

\begin{figure*}[t]
  \centering
  \includegraphics{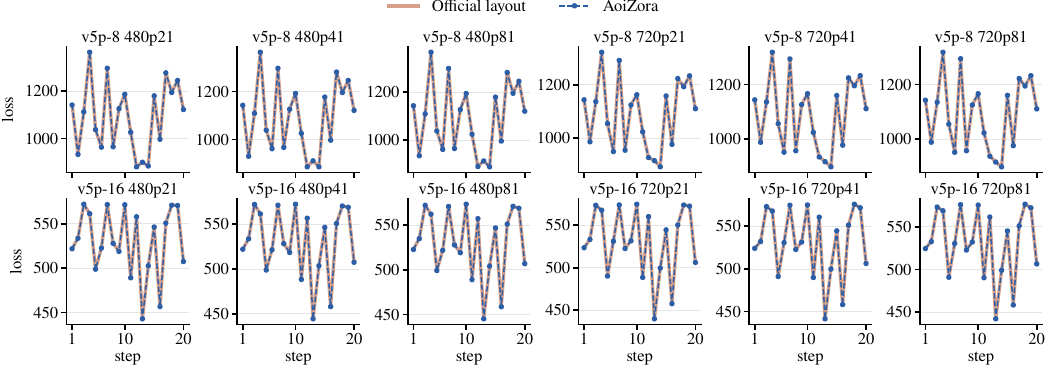}%
  \caption{Loss of Wan~2.1 1.3B over 20 training steps from the same initialization on the same data, under the official layout and \sys{}'s plan. Steps use different batches and diffusion timesteps, so the loss varies by step; the two sequences coincide.}
  \label{fig:app-loss}
\end{figure*}

\begin{table}[t]
  \centering
  \small
  \caption{Relative L2 distance between the 20-step loss sequences of \sys{}'s plan and the official layout, in units of $10^{-4}$.}
  \label{tab:app-loss}
  \vskip 0.1in
  \setlength{\tabcolsep}{2pt}
  \begin{tabular*}{\columnwidth}{@{\extracolsep{\fill}}lcc@{}}
    \toprule
    Shape & v5p-8 & v5p-16 \\
    \midrule
    480p21 & 2.43 & 2.52 \\
    480p41 & 2.40 & 2.43 \\
    480p81 & 2.34 & 2.52 \\
    720p21 & 2.46 & 2.49 \\
    720p41 & 2.38 & 2.43 \\
    720p81 & 2.51 & 2.52 \\
    \bottomrule
  \end{tabular*}
\end{table}

\subsection{PyTorch on GPUs}
\label{sec:eval-portability}

\begin{figure}[t]
  \centering
  \includegraphics{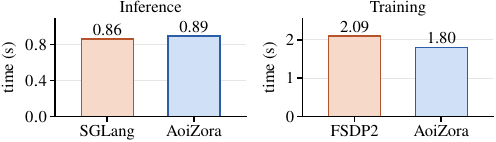}%
  \caption{Wan~2.1 14B on four H100 NVL GPUs: time of one denoising step at 480p41, and of one training step with two latent frames and global batch four.}
  \label{fig:gpu}
\end{figure}

\sys{} comes within 4\% of SGLang Diffusion~\cite{sglang} for inference and trains 1.16$\times$ faster than FSDP2~\cite{fsdp} (Figure~\ref{fig:gpu}), with the same planner core as on TPUs: only the importer, the realizer, the evidence adapter, and a GPU resource model belong to this stack. We run it on four H100 NVL PCIe GPUs, connected in pairs by NVLink bridges; the two pairs communicate over PCIe through the host, and there is no NVSwitch. We use PyTorch 2.14.1 and NCCL 2.30~\cite{nccl} and calibrate the resource model on microbenchmarks only. For inference, SGLang chooses its own parallel configuration, and both systems use the same cuDNN~\cite{cudnn} attention kernel and synchronized step timing, so \sys{} matches a system whose parallel strategies are written for this model family while planning from the unsharded model.

\subsection{Model Fidelity and Placement Ablation}
\label{sec:eval-ablation}

\begin{table}[t]
  \centering
  \small
  \caption{Ranking compiled pools of Wan~2.1 on v5e-8: Spearman's $\rho$ with measured step time (average ranks for ties) and regret, the selected plan's measured slowdown against the pool's fastest.}
  \label{tab:fidelity}
  \vskip 0.1in
  \setlength{\tabcolsep}{2pt}
  \begin{tabular*}{\columnwidth}{@{\extracolsep{\fill}}lccccc@{}}
    \toprule
     & & \multicolumn{2}{c}{Logical cost} & \multicolumn{2}{c}{Compiled} \\
    \cmidrule(lr){3-4} \cmidrule(l){5-6}
    Workload & Pool & $\rho$ & Regret & $\rho$ & Regret \\
    \midrule
    480p21 & 6 & 0.03 & 9.8\% & 0.83 & 0\% \\
    480p41 & 6 & $-0.51$ & 38.6\% & 0.89 & 0\% \\
    480p81 & 5 & $-0.35$ & 10.0\% & 0.90 & 0\% \\
    \bottomrule
  \end{tabular*}
\end{table}

\begin{table}[t]
  \centering
  \small
  \caption{Placement ablation of HunyuanVideo 13B inference at 544p on \mbox{v5e-16}: seconds per step of the plan each search stage selects. Rows 2 and 3 add one stage of the search each, and the last row disables placement search only.}
  \label{tab:ablation}
  \vskip 0.1in
  \setlength{\tabcolsep}{2pt}
  \begin{tabular*}{\columnwidth}{@{\extracolsep{\fill}}lcccc@{}}
    \toprule
    Plan & Mesh & 544p33 & 544p65 & 544p129 \\
    \midrule
    Integer program, unplaced & $2{\times}8$ & 0.474 & 1.344 & 2.658 \\
    + placement search & $2{\times}8$ & 0.410 & 1.203 & 2.321 \\
    + refinement (\sys{}) & $4{\times}4$ & 0.335 & 0.832 & 2.194 \\
    Refinement, no placement & $2{\times}8$ & 0.392 & 0.909 & 2.404 \\
    \bottomrule
  \end{tabular*}
\end{table}

\textbf{Compiled evidence, not the network model, corrects the ranking.} To separate what compiled evidence adds from the network model, we rank compiled pools two ways (Table~\ref{tab:fidelity}): by the integer program's logical cost, which prices the planner's own per-operation events with the same network model and axis-aligned assignments but sums them without the compiled schedule, and by the compiled prediction. Unlike \sys{}'s unplaced proposals, here the integer program proposes on every axis-aligned assignment; placement search and refinement are disabled, and every candidate runs on the assignment it was proposed on and is measured. The logical cost selects a $2{\times}4$ plan 9.8\% to 38.6\% slower than the fastest in all three pools; the compiled prediction selects the fastest, a one-dimensional plan, in all three, as Observation~1 anticipates. In the complete pools of the 36 full searches of \S\ref{sec:eval-inference}, with placement search and refinement enabled, the compiled prediction selects the fastest in 22 and stays within 0.74\% in 12 others; rebinding and recompilation priced every tested plan and placement pair identically (Table~\ref{tab:rebind}).

\begin{table}[t]
  \centering
  \small
  \caption{Median relative error of the resource model on held-out microbenchmark samples (p90 in parentheses).}
  \label{tab:app-heldout}
  \vskip 0.1in
  \setlength{\tabcolsep}{2pt}
  \begin{tabular*}{\columnwidth}{@{\extracolsep{\fill}}lccc@{}}
    \toprule
     & v5e-4 & v5e-8 & v5e-16 \\
    \midrule
    Collectives & 6.0\% (16.2\%) & 5.2\% (13.0\%) & 4.5\% (12.5\%) \\
    Point-to-point & 7.0\% & 4.4\% & 6.7\% \\
    Routing probes & \textendash & 3.2\% & 2.4\% \\
    Attention & \textendash & 6.5\% & 6.5\% \\
    Matrix products & \textendash & 4.8\% & \textendash \\
    \bottomrule
  \end{tabular*}
\end{table}

\begin{table}[t]
  \centering
  \small
  \caption{Median relative error of the resource model on isolated measurements of the collectives in compiled Wan~2.1 programs on v5e-8.}
  \label{tab:app-compiled}
  \vskip 0.1in
  \setlength{\tabcolsep}{2pt}
  \begin{tabular*}{\columnwidth}{@{\extracolsep{\fill}}lc@{}}
    \toprule
    Collective & Median error \\
    \midrule
    All-reduce & 7\% to 10\% \\
    All-gather & 6\% \\
    Collective permute & 0.7\% \\
    All-to-all & 3\% to 9\% \\
    \bottomrule
  \end{tabular*}
\end{table}

\begin{table}[t]
  \centering
  \small
  \caption{Median relative error on link-sharing samples held out of the fit, and the underestimate of a model that ignores link sharing.}
  \label{tab:app-sharing}
  \vskip 0.1in
  \setlength{\tabcolsep}{2pt}
  \begin{tabular*}{\columnwidth}{@{\extracolsep{\fill}}llcc@{}}
    \toprule
    Slice & Held-out case & Model & Sharing ignored \\
    \midrule
    v5e-8 & Groups of one collective & 5.6\% & 47\% \\
    v5e-8 & Transfers over one link & 1.0\% & 67\% \\
    v5e-16 & Strided groups & 6.5\% & 46\% \\
    v5e-16 & Transfers over one link & 5.0\% & 65\% \\
    \bottomrule
  \end{tabular*}
\end{table}

\textbf{The resource model is accurate in isolation and under contention.} The constants of each slice's resource model are fitted on the microbenchmarks of \S\ref{sec:network}, with every other payload size held out of the fit. Table~\ref{tab:app-heldout} lists the median relative error on the held-out samples, and Table~\ref{tab:app-compiled} the error on isolated measurements of the collectives in compiled Wan~2.1 programs. Link sharing is not fitted per scenario. To test this, the model is refitted without any sample in which two groups or two transfers share a link, and it predicts those samples from routes and capacities alone (Table~\ref{tab:app-sharing}); a model that ignores link sharing underestimates their time by 46\% to 67\%.

\textbf{Placement changes which sharding is selected.} On Wan 480p41 on v5e-16, a $4{\times}4$ sub-slice without wrap-around links, the integer program's best $2{\times}8$ sharding measures 0.824\,s in device-id order and 0.730\,s where placement search puts it, which pairs neighboring chips along the two-way axis. Refinement on that assignment stalls; restarted on the other axis-aligned assignment, which lays the eight-way axis on a $2{\times}4$ block, it proposes a different $2{\times}8$ sharding that measures 0.667\,s, 1.09$\times$ faster. This plan runs 10.4\% slower on the first assignment (11.8\% predicted): the best sharding depends on its placement (Observation~3).

\textbf{Placement search and refinement change the mesh together.} Table~\ref{tab:ablation} traces HunyuanVideo at 544p on v5e-16 stage by stage. Solved once per mesh without a device assignment, the integer program selects a $2{\times}8$ mesh in device-id order. Placement search moves it to a better assignment, 1.12$\times$ to 1.16$\times$ faster, and also places the $4{\times}4$ proposals, laying one four-way axis on $2{\times}2$ blocks. Refinement re-solves the $4{\times}4$ mesh on that assignment and selects it, another 1.06$\times$ to 1.45$\times$ faster. Without placement search, refinement re-solves the $4{\times}4$ mesh in device-id order and keeps the $2{\times}8$ mesh, 1.09$\times$ to 1.17$\times$ slower than \sys{}.

\subsection{Numerical Correctness}
\label{sec:eval-numerics}

\textbf{Inference.} Each plan runs one denoising step on synthetic weights and inputs, and its output is compared with the same step computed unsharded in FP32. As a precision floor, the unsharded step is also run in BF16, the precision the plans compute in. A plan's relative L2 error is 0.95 to 1.16 times the floor, so every plan stays within BF16 precision; some fall slightly below the floor because sharding changes the order of reductions.

\textbf{Training.} Each selected training plan and the official layout run 20 training steps from the same initialization on the same data. Figure~\ref{fig:app-loss} plots both loss sequences, and Table~\ref{tab:app-loss} lists their relative L2 distance, $2.3\times10^{-4}$ to $2.5\times10^{-4}$ in all 12 settings, below the unit roundoff of BF16 ($2^{-8} \approx 3.9\times10^{-3}$).

The checks of \S\ref{sec:evidence} cover the plan graph and realization before any plan runs.

\section{Related Work}
\label{sec:related}

\textbf{Auto-parallel planning and DiT parallelism.} Planners such as FlexFlow~\cite{flexflow}, Alpa~\cite{alpa}, Unity~\cite{unity}, Galvatron~\cite{galvatron}, nnScaler~\cite{nnscaler}, and Mist~\cite{mist} search sharding strategies over a logical graph and price them with analytical models or operator profiles, and Automap and PartIR~\cite{automap,partir} search partitioning tactics in a compiler IR before code generation. They leave the device order to the framework's default mesh; NEST~\cite{nest} places pipeline stages and parallel groups on the levels of hierarchical networks with a dynamic program over profiled and simulated costs; it abstracts a torus into hop-distance classes and does not choose the chip of each rank or see the compiled program. For DiTs, experts fix the strategy per model: sequence parallelism with Ulysses and ring attention~\cite{usp,xdit}, patch parallelism~\cite{distrifusion,pipefusion}, and the context-parallel configurations of MaxDiffusion and SGLang~\cite{maxdiffusion,sglang}. \sys{} instead prices plans by their compiled programs on a calibrated network model, chooses the chip of every logical rank, and competes with the DiT expert configurations without a model-specific recipe.

\textbf{Performance models of compiled programs.} Learned cost models predict the run time of compiled tensor programs on one TPU~\cite{kaufman,tpugraphs}, and simulators such as Daydream, dPRO, Lumos, and ASTRA-sim~\cite{daydream,dpro,lumos,astrasim} replay the trace of an executed configuration, or a modeled workload, on a network model. The former stop at one device; the latter need an execution or a workload description whose placement is already fixed. \sys{} extracts evidence from the compiler without execution and records collectives over logical ranks, so one compilation is priced under many placements.

\textbf{Topology mapping.} Mapping communicating tasks onto tori and meshes is a mature HPC problem~\cite{bokhari,agarwal,hoefler,bhatele,rubik}: given an application's communication graph, mappers minimize hop-bytes or link congestion, and Rubik maps collectives over sub-communicators by structured transformations of rank blocks. On TPUs, JAX maps logical mesh axes to torus axes by fixed rules~\cite{jax}, layouts for the TPU v4 torus are designed by hand~\cite{pope}, and TopoOpt and optical reconfiguration change the topology rather than the program~\cite{topoopt,tpuv4}. All take the communication pattern as input, whereas in a DiT plan the sharding chooses it. \sys{} starts from structured assignments in the spirit of these mappers, but prices each on its compiled program and searches it jointly with the sharding.

\section{Conclusion}
\label{sec:conclusion}

We presented \sys{}, a compiler-agnostic auto-parallelization framework that chooses the logical mesh, the layout of every tensor and operation, and the device assignment of a DiT by pricing the programs its compiler emits, without executing any candidate. Its core is evidence, a compiler-neutral record of a compiled program over logical ranks that a calibrated network model prices under any placement, so that the placement can influence which sharding is selected without compiling every combination. \sys{} reduces DiT inference latency on TPU v5e by up to 2.29$\times$ over expert and planner baselines and speeds up training on TPUs and GPUs.

\section*{AI Usage Disclosure}
We used Claude (Anthropic), an AI assistant, under our direction. For writing, it helped draft, edit, and polish the text and check the consistency of notation, references, and cross-references. For experiments, it helped write scripts that run experiments, collect data, and plot figures, and it ran experiments on our TPU allocations. The research problem, ideas, system design, choice of experiments, analysis, and conclusions are our own; we reviewed every change and result it produced and take full responsibility for the content of this paper. No content of the paper was generated by an AI tool without our review.

{\small
\bibliographystyle{plain}
\bibliography{references}
}

\end{document}